\documentclass[12pt]{article}
\newcommand{\be}{\begin{equation}} 
\newcommand{\ee}{\end{equation}}
\newcommand{\bea}{\begin{eqnarray}} 
\newcommand{\eea}{\end{eqnarray}}

\usepackage{graphicx,amssymb}

\def\IR{\relax{\rm I\kern-.18em R}}
\def\IN{\relax{\rm I\kern-.18em N}}

\def\Ker{{\rm Ker}\,}

\def\e{\epsilon}
\def\z{\zeta}

\font\cmss=cmss10 \font\cmsss=cmss10 at 7pt
\def\IZ{\relax\ifmmode\mathchoice
{\hbox{\cmss Z\kern-.4em Z}}{\hbox{\cmss Z\kern-.4em Z}}
{\lower.9pt\hbox{\cmsss Z\kern-.4em Z}}
{\lower1.2pt\hbox{\cmsss Z\kern-.4em Z}}\else{\cmss Z\kern-.4em Z}\fi}

\begin{document}

\title{Relative entropy for compact Riemann surfaces} 
\author{J. Gaite
\\ 
{\it Instituto de Matem{\'a}ticas y F{\'\i}sica Fundamental, CSIC,}\\ 
       {\it Serrano 123, 28006 Madrid, Spain}\\
and\\
{\it Laboratorio de Astrof{\'\i}sica Espacial y F{\'\i}sica Fundamental,}\\
{\it Apartado 50727, 28080 Madrid, Spain}
}
\date{July 29, 1999}

%
\maketitle

\begin{abstract}
  The relative entropy of the massive free bosonic field theory is
  studied on various compact Riemann surfaces as a universal quantity
  with physical significance, in particular, for gravitational
  phenomena. The exact expression for the sphere is obtained, as well
  as its asymptotic series for large mass and its Taylor series for
  small mass. One can also derive exact expressions for the torus but
  not for higher genus. However, the asymptotic behaviour for large
  mass can always be established---up to a constant---with heat-kernel
  methods.  It consists of an asymptotic series determined only by the
  curvature, hence common for homogeneous surfaces of genus higher
  than one, and exponentially vanishing corrections whose form is
  determined by the concrete topology. The coefficient of the
  logarithmic term in this series gives the conformal anomaly.
\end{abstract}
\global\parskip 3pt \vskip .5cm {\small
  }
\newpage

\section{Introduction}

The entropy of a statistical model relative to its critical point has
been shown to be an interesting quantity in field theory, especially
in regard to the renormalization group \cite{I-OC}. On the one hand,
it exhibits better behaviour than the free energy when the ultraviolet
cutoff is sent to infinity and, on the other hand, it is monotonic
with the coupling constants, unlike the free energy. This second
property makes it suitable to embody the irreversible nature of the
renomalization group, which can in particular be substantiated in a
finite geometry as monotonicity with respect to its characteristic
scale \cite{I}. The computation of the {\em relative entropy} for various
models on a cylinder clearly shows its monotonicity \cite{I}. The
cylinder is appropriate to illustrate finite size-effects but it may
not be the finite geometry of choice in the context of
renormalization-group irreversibility. There is a celebrated result on
renormalization-group irreversibility in two-dimensional ($2d$) field
theories, Zamolodchikov $C$ theorem. The monotonicity theorem for the
relative entropy on the cylinder, once it is conveniently formulated,
resembles Zamolodchikov $C$ theorem \cite{I}.  However, this
resemblance can hardly lead to a direct relationship, since the proof
of Zamolodchikov $C$ theorem demands rotation as well as translation
symmetry. In other words, that proof demands to consider a maximally
symmetric space, namely, the sphere, the plane or the hyperbolic
plane. Both the sphere and the hyperbolic plane possess an infrared
scale, the curvature radius, but only the sphere is finite and
therefore the computation of the relative entropy on the sphere is of
particular value.

From a different point of view, the calculation of the spectrum of the
Laplacian operator on general compact Riemann surfaces has held the
interest of first mathematicians \cite{Kac,McKean} and second
physicist \cite{Alva,D'H-Ph} for some time.  The partition function of
the bosonic massive field theory on a compact Riemann surface is a
global object which can be constructed from the knowledge of the
spectrum of the Laplacian operator. Since this spectrum is highly
dependent on the topologic and geometric properties of the Riemann
surface, these properties are reflected by the partition function.
However, this function, or say the free energy, is ultraviolet (UV)
divergent and hence ill defined.  Fortunately, the relative entropy of
the $2d$ bosonic massive field theory is UV finite \cite{I} and so it
is likely to have a r\^ole in the geometrical characterization of a
Riemann surface. This characterization consists of local parameters,
namely, the curvature, and global parameters, specifying the boundary
conditions and topologically significant. The study of homogeneous
surfaces will provide insight into the dependence of the relative
entropy on both types of parameters.

Generically, the relative entropy is not related to the quantum field
theory entropy, but there is a direct relation on the torus (or
cylinder) geometry.  The relative entropy is a {\em geometric
  entropy}, of the sort already considered in connection with the
entropy of black holes \cite{geoment}. Indeed, the geometry relevant
to this case is that of the cone, which is non-compact and is actually
related to the cylinder geometry, and hence to the usual quantum field
theory entropy, as were analyzed in Ref.~\cite{I}. One can expect that
the relative entropy for homogeneous spaces will be applicable in a
cosmological context, once suitably generalized.  Thus the results of
this paper must have some bearing on entropic considerations in {\em
  de Sitter} and {\em anti-de Sitter} space-time.  An attempt at
introducing the maximum entropy principle in quantum cosmology has
been made in Ref.~\cite{Rosa}. On the other hand, the application of
scaling and renormalization group concepts in gravitation
\cite{gravRG} and cosmology \cite{cosmoRG} is gaining momentum.
Therefore, it seems interesting to study of properties of the entropy
relative to the scales defining some curved space.  Furthermore, the
r\^ole of the relative entropy as a function monotonic with the
renormalization group may as well have some relevance in modern
theories of quantum gravity, as recent work seems to indicate
\cite{AdSRG}.

Therefore, our main concern here will be the computation of the
universal relative entropy of the $2d$ bosonic massive field theory on
homogeneous and compact Riemann surfaces, for its own sake and with a
view to its application in connection with Zamolodchikov $C$ theorem.
For the plane and cylinder, the relative entropy has been computed in
Ref.~\cite{I}.  Since we are now concerned with compact Riemann
surfaces, we shall first focus on the simplest case, namely, the
sphere.  In general, the topological classification of compact Riemann
surfaces is given by their {\em genus}, that is, the number of handles
in a three-dimensional embedding. The genus of the sphere is zero, of
course. Compact Riemann surfaces of higher genus, with zero or
negative constant curvature, will also be considered here, even though
they are not globally isotropic, since they are derived from the plane
or the hyperbolic space by imposing boundary conditions on a finite
domain which break rotation invariance. The case of zero curvature is
the torus and is actually related to the cylinder, treated in
Ref.~\cite{I}.  Compact Riemann surfaces of genus $g>1$ are always
related to the hyperbolic plane.  The spectrum of the Laplacian on
them is too complicated to allow derivation of closed expressions for
the relative entropy. Thus, the case $g>1$ will be discussed summarily
and only some general properties of the free energy and the relative
entropy will be extracted. The method used to obtain these properties,
namely, the heat-kernel method, is however of general interest and we
shall dedicate considerable attention to it.

A large literature has been devoted to the computation of vacuum
energy densities on various manifolds, mostly in regard to field
theory in curved space-time and to the Casimir effect. This energy is
divergent, of course, and needs regularization.  The usual technique
is the zeta-function regularization \cite{Dowker,Matt,zeta}, as
introduced earlier in the mathematical literature \cite{McKean}. It is
related to the heat-kernel representation that will be utilized here
by a Mellin transform. We will see how this relationship materializes
for $g>1$ Riemann surfaces in the last section.\footnote{For a
  comprehensive review of all these techniques, see
  Ref.~\cite{Bytetal}.}  However, since the relative entropy is a
universal quantity we do not need to bother with prescribing any
regularization method and we shall only do it to make connections, for
instance, with the conformal anomaly or with partition functions in
string theory.

\section{The relative entropy of the sphere}

The free energy $W$, namely, minus the logarithm of the partition
function, for the cutoff bosonic massive field theory 
can be expressed as \cite{I-OC,I}
\begin{equation}
W[m,\Lambda] \equiv -\ln Z[m,\Lambda] =
{1\over2}\sum_{{\vec p}}\ln{{{\vec p}}^{\,2}+m^2\over\Lambda^2}.
\label{W}
\end{equation}
The set of momenta to be summed depends on the type of geometry and is
such that ${{\vec p}}^{\,2} < \Lambda^2$. If we try to remove the
cutoff we see that $W$ is UV divergent, and hence it is non universal.
The eigenvalues of the Laplacian on the sphere are well known, namely,
${\vec p}^{\,2} = l\,(l+1)/R^2$, where $R$ is the sphere radius. Thus,
the sum over momenta is a sum over $l$ and we write
\begin{equation}
W[r,\Lambda] =
{1\over2}\sum_{l=1}^{l_{\rm max}}(2\,l +1)\,\ln{l\,(l+1)+r\over(\Lambda\,R)^2},
\label{Ws}
\end{equation}
where we have introduced the dimensionless coupling $r = (m\,R)^2$ and
the UV cutoff is related to the maximum value of $l$, $\Lambda \approx
l_{\rm max}/R$. We have also removed the zero mode $l=0$ from the sum,
which is not allowed for $r=0$, and we have taken the degeneracy into
account by the factor $2\,l +1$.

From the previous expression of $W$ one can obtain the relative
entropy as a universal quantity, that is, as a convergent series in
the limit $l_{\rm max} \rightarrow \infty$. However, the sum of that series
is very hard to carry out, so we choose another way. One can lower the degree
of divergence of $W$ by taking derivatives with respect to $r$. In
fact,
\be
{dW \over dr} = {1\over2}\sum_{l=1}^{l_{\rm max}} \frac{2\,l +1}{l\,(l+1)+r},
\ee
which is still logarithmically divergent. We can remove this
divergence just by substracting its value at $r=0$ and write, in the
limit $l_{\rm max} \rightarrow \infty$,
\be
{dW \over dr} - \left.{dW \over dr}\right|_{r=0} = 
-{r\over2}\sum_{l=1}^{\infty} \frac{2\,l +1}{[l\,(l+1)+r]\,l\,(l+1)}.
\ee
Hence, we define the function $U(r) := (dW/dr) - (dW/dr)_{r=0}$, which 
turns out to be computable in terms of the digamma function $\psi(x)$.
The full expression is rather long; it is given in an appendix. 

The function $U(r)$ has interest on its own, since it is the substracted 
energy, but we are more interested in the relative entropy. This can be 
expressed in terms of $U(r)$ as
\be
S(r) = W(r) - W(0) - r\,{dW(r)\over dr} =\int\limits_0^r U(s)\, ds - r\,U(r).
\ee
Unfortunately, this integral cannot be done in closed form. However, it is 
possible to establish the behaviour of $S(r)$ for small or large $r$. For 
large $r$ the correlation length is much smaller than the radius of the 
sphere and the result in the plane, $S(r)=r/(8\pi)$, should be relevant. 
As $R \rightarrow \infty$ the sum over $l$ can be substituted by an integral, 
$$
\sum_{l=1}^{\infty}(2\,l +1) \rightarrow 
A\int\limits_0^\infty \frac{d^2p}{(2\pi)^2},
$$
where the area of the sphere is $A= 4\pi R^2$. We see that to compare
with the value in the plane we must multiply this value by $4\pi$. To extract 
the dominant large $r$ behaviour of $U(r)$, we use $\psi(x) \approx \ln x$. 
A lengthy but straightforward calculation shows that $U(r) \approx -\ln r/2$. 
Hence, using first an integration by parts, 
$$
S(r) = -\int\limits_0^r s\,U'(s)\,ds \approx {r\over2},
$$
as expected.

The full asymptotic expansion of $S(r)$ near infinity results from that
of $U(r)$, which in turn can be worked out with help of the known
asymptotic expansion of $\psi(x)$. However, this procedure yields the
asymptotic expansion of $S(r)$ only up to a constant, because the
condition $S(0)=0$ cannot be implemented. It begins as \be S(r)
\approx {r\over2} - {\frac{\ln r}{3}} - {\frac{1}{15\,r}} -
{\frac{2}{105\,{r^2}}} - {\frac{4}{315\,{r^3}}} +
{{O}({\frac{1}{r}})}^{4}.  \label{Ssa} \ee The presence of a
subleading term $\ln r$ was to be expected, owing to the existence of
a conformal anomaly for any Riemann surface, coming from the
logarithmic divergence of the critical free energy. 
The logarithmic term
of the free energy turns out to be
\be 
W = -\frac{\chi}{12}\,\ln ({\Lambda^2\,R^2}),
\label{critW} 
\ee 
where $\Lambda$ is the UV cutoff, $R$ the size, and $\chi$ is the
integral of the curvature divided by $4\pi$, equal to the
Euler-Poincar\'e number of the Riemann surface, according to the
Gauss-Bonnet theorem. The calculation of the conformal anomaly from
the logarithmic divergence of the critical free energy is probably
very old but was popularized by the development of string theory
\cite{Alva}.  It was further discussed in Ref.~\cite{Matt}, in the
context of the Casimir energy. In the presence of mass, a term
proportional to $\ln (m^2/\Lambda^2)$ must appear when the correlation
length $m^{-1}$ becomes smaller than $R$. Hence, it is not surprising
to have the term $\ln r$ in the previous expansion.  The concrete way
in which it appears will be explained in section 4, when we consider
the heat-kernel derivation of the asymptotic expansion. This is a much
more effective method to find the large $r$ behaviour, capable of
providing the generic form of the coefficients of the asymptotic
series for arbitrary Riemann surfaces.

A plot of $S(r)$ is shown in Fig.\ 1, in comparison with the
asymptotic behaviour given by just the two growing terms in the
previous formula (\ref{Ssa}). The agrement is quite remarkable, even
almost down to $r=1$. The numerical value of the missing constant is
approximately $0.254381$.

\vspace{1cm}
\includegraphics[width=7cm]{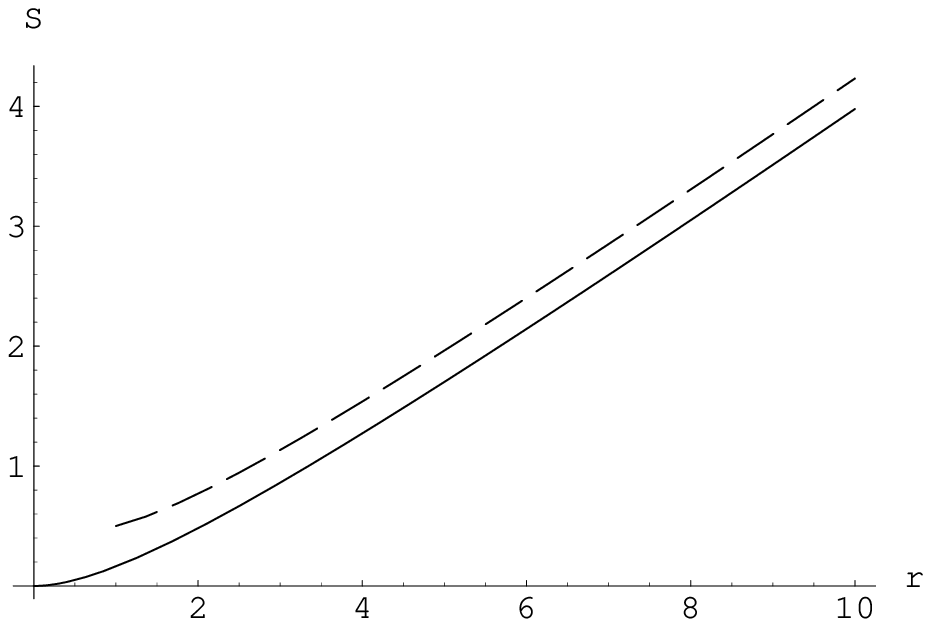}
\hspace{0.5cm}
\includegraphics[width=7cm]{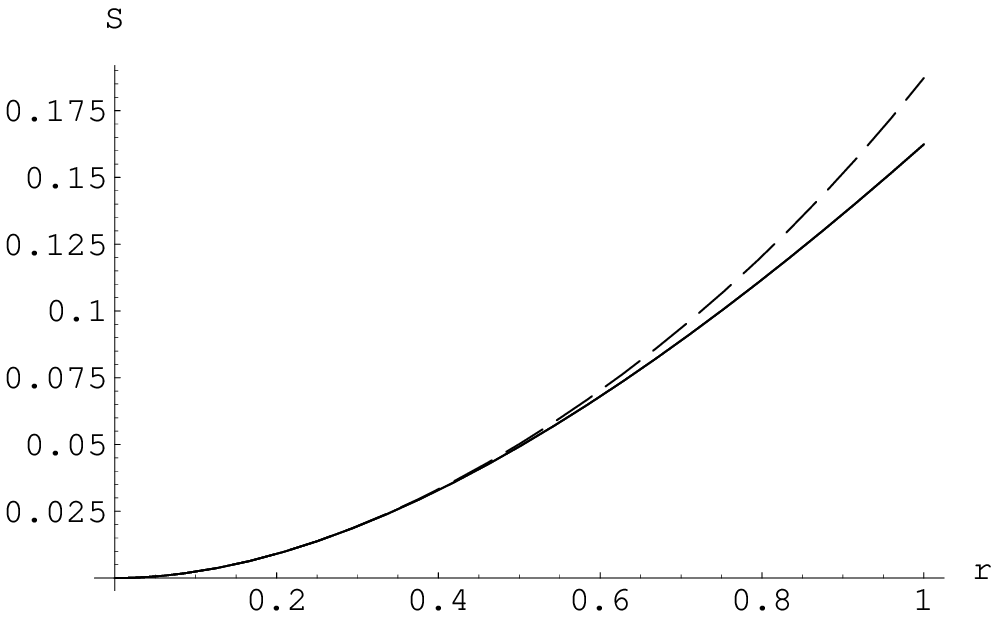}
\vspace{0.5cm}

\centerline{Fig.\ 1: $S(r)$ on the sphere compared to the first two
  terms of}
\centerline{the asymptotic formula (\ref{Ssa}) and to the Taylor
  expansion (\ref{SsT})} \vspace{1cm}
\noindent
The small $r$ behaviour of $S(r)$ is given by the power series
expansion near $r=0$, which is easy to derive since $S'(r) = -r\,U'(r)$.
We obtain that
\be
S(r) = \frac{1}{4}\,r^2 + \frac{1}{3}\,[2+\psi^{(2)}(1)]\,r^3 + 
\frac{3}{16}\,[12+ 5\,\psi^{(2)}(1) - \psi^{(2)}(2)]\,r^4 + {{O}(r)}^{5},
\label{SsT}
\ee
where $\psi^{(n)}(x)$ is the polygamma function. The behaviour provided 
by the series truncated to this order is compared with the total $S(r)$ in 
the second plot of Fig.~1. 
The radius of convergence of the Taylor series is determined by the singular
points of $U(r)$. Since the only singularities of $\psi(x)$ are simple
poles for non-positive integers, it is easy to see that the
singularity closest to $r=0$ is a simple pole at $r=-2$, and hence 
the radius of convergence is 2.

\section{The relative entropy of the torus}

We consider the torus as a rectangular box with periodic boundary
conditions, which is the natural finite geometry in many applications.
One can slightly generalize the boundary conditions by considering
periodicity along two non-orthogonal directions, that is, by letting
the box be a parallelogram. Although a parallelogram is in principle
equivalent to a rectangle by an affine transformation, this is only
true in real geometry, because in complex geometry such transformation
is not allowed. Nevertheless, we shall consider a rectangle for
simplicity and deduce the more general form from holomorphic
factorization.

The partition function on the torus in the critical theory, $m=0$, is
essentially the modulus of Dedekind's function $\eta(\tau)$. The
classical proof of this result involves the use of the proper time
representation and Poisson resummation, after analytical continuation
in the manner of $\zeta$-function regularization, since the partition
function is UV divergent. This method can be extended to the
non-critical theory \cite{ItDr} (also \cite{NaOC}). However, we favour
a method similar to the one used for the sphere, where we calculate
the energy $U(r)$. The substraction of the critical value will not be
necessary since the UV divergence lends itself to straightforward
identification.

Let us $L$ and $M$ denote the periods in the horizontal and vertical
directions, respectively. Then (see appendix)
\be
W(r) = \frac{M}{2}\sum_{l=-\infty}^{\infty}\e(l)+
\sum_{l=-\infty}^{\infty}\ln\left[1- e^{-M\,\epsilon(l)}\right] + C,
\ee
where we have introduced the one-boson energies $\e(l) =
\sqrt{(2\pi\,l/L)^2+m^2}$ and where $C$ is a constant irrelevant for
the relative entropy.

Now, we may notice that the previous expression for $W$ can be
interpreted as the free energy of quantum $1d$ bosons confined in a
segment of length $L$ at finite temperature $T = 1/M$ and constitutes
a slight generalization of the expression for the cylinder considered
before \cite{I}.  To prevent a divergence when $r=0$ we remove the
zero mode $l=0$ and write
\be
W(r) = M\sum_{l=1}^{\infty}\e(l)+
2\sum_{l=1}^{\infty}\ln\left[1- e^{-M\,\epsilon(l)}\right]. 
\label{Wte}
\ee
The UV divergences concentrate in the $1d$ vacuum energy $E_0
=\sum_{l=1}^{\infty}\e(l)$. However, for $r=0$ they can be
$\z$-regularized to give $E_0 = (2\pi/L)\,\zeta(-1) =
-\pi/(6\,L)$. The other term is the free energy of the bosonic
excitations of the vacuum and is finite. For $r=0$ it combines with
$E_0$ to yield $W \equiv -\ln Z = \ln \eta(q)^2$, where $\eta(q)$ is
the Dedekind function, $q=\exp 2\pi i\tau$ and the modular parameter
is $\tau = i\,(M/L)$. In the case of a parallelogram the modular
parameter is, of course, complex and $Z = 1/[\eta(q)\,\eta({\bar
  q})]$.

Most of the discussion about the relative entropy of the cylinder in
Ref.~\cite{I} holds as well for the torus.  Hence, we change the notation
for the vertical period, $M \rightarrow \beta$, in accord with the
$1d$ thermodynamic interpretation.  The specific relative entropy is
related to the quantum $1d$ entropy of free bosons, $S_q$, as
\be
S(r;L,\beta) = S(r;L)
 - {S_q\over 2\,L\,\beta} + {\pi\over 6\,\beta^2},   
\label{StoSG}
\ee
where
\be
S(r;L) = e_0(r;L) - e_0(0;L) - r\,{de_0(r;L)\over dr}, 
\quad e_0(r;L) := {E_0(r;L)\over L}.
\ee
The quantity $e_0(0;L)= -{\pi/(6\,L^2)}$ is dual to
$-{\pi/(6\,\beta^2)}$.  Note that $$S(r;L) =
\lim_{\beta\rightarrow\infty} S(r;L,\beta)$$
is the relative entropy
of the cylinder. Of course, $\lim_{L\rightarrow\infty} S(r;L) =
\pi\,{r/2}$.  While $S(r;L,\beta)$ is modular invariant, the $1d$
quantum entropy $S_q$ comes only from the free energy of the bosonic
excitations, namely, the second term in (\ref{Wte}), and is not
modular invariant.  $S(r;L,\beta)$ can be computed by the heat-kernel
method, studied in next section, in terms of a double series of Bessel
functions. Its cylinder limit coincides with the single series of
Bessel functions computed in Ref.\ \cite{I}.  On the other hand, it is
feasible to obtain the perturbative expansion of $W$ near $m=0$
\cite{ItDr}, and from it the expansion of $S(r;L,\beta)$. The
asymptotic expansion for large $m$ is just given by the value of $S$
for the plane and it comes from the first term, $S(r;L)$, in
(\ref{StoSG}), since the second one decays exponentially with $L$. A
more precise description of this asymptotic behaviour is provided by
the heat-kernel method.

The quantum $1d$ entropy of free bosons, $S_q$, can be related to the
entanglement entropy of black holes in $1+1$ dimensions. The Euclidean
version of {\em Rindler space} is just the punctured plane, which is
conformally diffeomorphic to the cylinder. In addition, the
entanglement entropy of this space can be computed with a path
integral that reduces it to an ordinary entropy, when considered on
the cylinder \cite{geoment}. Therefore, one can use the formulas
above, taking into account that the conformal transformation implies
that the modular parameter becomes a function of the quotient of the
two cutoffs for the radius, namely, $i/\tau = L/\beta = \ln
(r_1/r_0)$. The complete expression of the cylinder entropy can be
deduced from the results in Ref.~\cite{I}. The first term of the low
mass expansion agrees with the one calculated by other methods in
Ref.~\cite{geoment}.

\section{Heat-kernel techniques}

The computation of thermodynamic quantities for the sphere or the
torus, regardless of ulterior difficulties, begins with the
preliminary step of determining the eigenvalues of the Laplacian.
Unfortunately, even this preliminary step cannot be taken for an
arbitrary compact Riemman surface. Therefore, one cannot start from a
more or less formal but explicit expression of those quantities.
Nevertheless, the situation is not hopeless for there is a powerful
method to extract information on thermodynamic quantities, namely, the
heat kernel representation of Green functions
\cite{Kac,McKean,Alva,D'H-Ph}. We now study this method, for the
purpose of applying it to $g>1$ Riemman surfaces and for its own sake:
actually, it provides useful insight for the sphere and the torus.

First, let us recall that
\be
W(m) = -{1\over2}\,Tr_{{\vec x},{\vec y}}\,\ln G({{\vec x},{\vec y}}),
\quad U(m) = {1\over2}\,Tr_{{\vec x},{\vec y}}\,G({{\vec x},{\vec y}}),
\ee
with $G({{\vec x},{\vec y}})$ the Green function of the Helmholtz equation,
namely, $(\Delta + m^2)\,G({{\vec x},{\vec y}})=\delta({{\vec x},{\vec y}})$. 
If we introduce the Green function of 
the heat equation
\be
\Delta\,\phi= {\partial\phi\over\partial t},
\ee
called the heat kernel, $K({{\vec x},{\vec y}};t)$, we can express $G$ as
a Laplace transform,
\be
G_{r}({{\vec x},{\vec y}}) = \int\limits_0^\infty dt\,
e^{-r\,t}\,K({{\vec x},{\vec y}};t),
\ee
after defining $r:=m^2$. Furthermore,
\be
U(r) = {1\over2}\int\limits_0^\infty dt\,
e^{-r\,t}\,Tr_{{\vec x},{\vec y}}\,K({{\vec x},{\vec y}};t), 
\label{Uhk}
\ee
and integrating over $r$,
\be
W(r) = -{1\over2}\int\limits_0^\infty {dt\over t}\,
e^{-r\,t}\,Tr_{{\vec x},{\vec y}}\,K({{\vec x},{\vec y}};t).
\label{Whk}
\ee
In the plane 
\be
K({{\vec x},{\vec y}};t)={1\over 4\pi t}\,
\exp-{|{{\vec x}-{\vec y}}|^2\over 4\,t},
\ee
so the specific energy is
\be
{U(r)\over A} = {1\over2}\int\limits_0^\infty {dt\over 4\pi t}\,e^{-r\,t},
\ee
with $A =\int d^2x$, the total area. To avoid the logarithmic UV 
divergence at $t =0$ we may take a further derivative,
\be
{U'(r)\over A} = -{1\over8\pi}\int\limits_0^\infty dt\,e^{-r\,t}
= -{1\over8\pi r}.
\ee
After integrating twice over $r$,
\be
{W(r)\over A} = -{1\over8\pi}\,(r\,\ln r + C_1\,r + C_2),
\ee
where $C_1$ and $C_2$ are two integration constants that are actually 
infinite due to UV divergences. Of course, this expression agrees with 
the one derived in \cite{I} using UV regularization in momentum space.

In a general curved surface $Tr_{{\vec x},{\vec y}}\,K({{\vec x},{\vec
    y}};t) = \int d^2x\, K({{\vec x},{\vec x}};t)$ and it is the integrated
heat kernel for coincident points. Note that $K({{\vec x},{\vec
    x}};t)$ explicitly depends on the point ${\vec x}$. However, for
homogeneous surfaces it becomes independent of it. According to
(\ref{Uhk}), 
\be {U(r)\over A} = {1\over2}\int\limits_0^\infty
dt\,e^{-r\,t}\,K(t),
\label{UK}
\ee
where $K(t)$ is the heat kernel for coincident points. 
If the eigenvalues of the Laplacian are available, 
$\Delta\,\phi_n = \gamma_n\,\phi_n$, the heat kernel is just
\be
K(t) = \sum_{n=0}^\infty e^{\gamma_n\,t}.    \label{Ksum}
\ee
The first eigenvalue is $\gamma_0 \equiv 0$, corresponding to the
constant solution; the others are negative and ordered by their
absolute value---in fact, $\gamma_1 < -1/4$ \cite{McKean}.
Substituting for $K(t)$ in the integral representation (\ref{UK}) we
recover the expression for the energy used before,
\be
{U(r)\over A} = {1\over2}\sum_{n=0}^\infty \frac{1}{-\gamma_n + r}.
\ee

For example, if we consider the torus, $\gamma_n = -(2\pi)^2\,|k|^2$, 
where $k \in \Omega^*$, the lattice dual to $\Omega$, 
the one definig periodicity on the torus. Then we can use 
the Jacobi identity to write
\be
K(t) = \sum_{k\in\Omega^*} \exp [-(2\pi)^2\,|k|^2\,t] 
= \frac{A}{4\pi t} \sum_{\omega\in\Omega} \exp [-{|\omega|^2\over 4\,t}].
\label{Jac}
\ee
The latter sum is $1 + O[\exp (-1/t)]$, reproducing the result 
for $U$ in the plane plus corrections vanishing exponentially when 
$r \rightarrow\infty$.  

Generically, the heat kernel $K(t)$ 
admits an asymptotic power series expansion at $t=0$ \cite{Mina},
\be
K(t) \approx {1\over 4\pi t} \left(1 + \sum_{n=1}^\infty a_n\, t^n\right),
\label{aK}
\ee
in terms of some numerical coefficients determined by the Riemann
curvature. This expansion, upon integration over $t$, leads to
asymptotic expansions of thermodynamic quantities for large $r$.  For
the hyperbolic plane, it can be calculated from the integral of
$\exp{-\nu^2\,t}$ times the density of states of frequency
$\nu$ \cite{OtteTa,Campo}:
\be 
K(t) = \frac{e^{t/4}}{2\pi}
\int\limits_0^\infty d\nu\, \nu\,{\tanh (\pi\nu)}\,{e^{-\nu^2\,t}}.
\label{iK}
\ee 
It yields
\be 
K(t) = \frac{e^{t/4}}{4\pi t}\left\{1 -
    \sum_{n=1}^\infty  \frac{1- 2^{1-2n}}{n!}\,B_{2n}\,t^n\right\},
\label{Ksa}
\ee
where $B_{2n}$ are the Bernoulli numbers.
The expansion for the sphere is related to it by analytic continuation
from the hyperbolic to the elliptic geometry and has an additional
$(-1)^n$ in the sum. 

From the expression (\ref{UK}) for the energy we can derive the heat kernel 
integral for the specific relative entropy,
\be
{S(r)\over A} = {1\over2}\int\limits_0^\infty {dt\over t}\,
[1-(1+r\,t)\,e^{-r\,t}]\,K(t),
\label{entK}
\ee
which can be easily shown to be convergent at $t=0$, taking into
account the behaviour of the heat kernel for small $t$ (\ref{aK}).
For a compact surface, it also converges as $t \rightarrow\infty$ if
we remove the zero mode, since then $K(t)$ decays exponentially,
according to Eq.\ (\ref{Ksum}).  To obtain the asymptotic $r
\rightarrow\infty$ expansion we could be tempted to just substitute
the aymptotic expansion (\ref{aK}) into this integral.  However, we
would encounter that the coefficients are given by integrals divergent
as $t \rightarrow\infty$. The problem is that we cannot prolong the
aymptotic expansion (\ref{aK}) to $t=\infty$. However, the aymptotic
expansion (\ref{aK}) can be substituted into the integral for $S'(r)$,
\be
{S'(r)\over A} = -r\,{U'(r)\over A} = 
{r\over2}\int\limits_0^\infty dt\,e^{-r\,t}\,t\,K(t),
\label{Sp}
\ee
to provide its asymptotic expansion $r \rightarrow\infty$, owing to
Watson's lemma \cite{BenOrs}.  This expansion can in turn be subjected
to indefinite integration to yield $S(r)$, of course, up to a
constant:
\be
{S(r)\over A} \approx -{1\over8\pi} 
\left(-r - a_1\,\ln r + \sum_{n=1}^\infty a_{n+1}\,
\frac{(n+1)!}{n}\,r^{-n}\right).
\ee
We can obtain the whole asymptotic series of $S(r)$ for the sphere by
just taking the coefficients from the corresponding asymptotic series
of $K(t)$, given by Eq.~(\ref{Ksa}) with $(-1)^n$ inserted. The
coefficients of its succesive terms coincide with the ones in the
previous expression (\ref{Ssa}) except for the one of the logarithmic
term.  This term, which was already remarked, is particularly
interesting and we now discuss it in more detail.

\subsection{The logarithmic term in the asymptotic
expansion of $S(r)$ and the conformal anomaly}

The logarithmic term in the asymptotic
expansion of $S(r)$ comes from the constant term in the asymptotic
power series of $K(t)$ (\ref{aK}), namely, $a_1/(4\pi)$.  In particular,
formula (\ref{Ksa}) yields $a_1 = \pm 1/3$, for the sphere and
hyperbolic plane, respectively. If we substitute this constant for
$K(t)$ in (\ref{Sp}) we obtain
\be
{S'(r)\over A}= -r\,{U'(r)\over A} = 
{r\over2}\int\limits_0^\infty dt\,e^{-r\,t}\,t\,{a_1\over 4\pi}  = 
{a_1\over8\pi\,r}.
\ee
(Notice the slight abuse of notation, for we deal here with truncated
quantities.)  For a curved compact homogeneous surface the
Gauss-Bonnet theorem implies that $A=2\pi\,|\chi| = 4\pi\,|g-1|$,
where $\chi$ is the Euler-Poincar\'e number.  Then the coefficient of
the logarithmic term of the total value of $S(r)$ is
$${a_1\,A\over 8\pi} = {\chi\over 12}= {1-g\over 6} ,$$ 
only related to the topology of
the Riemann surface. This result also holds for compact surfaces of
variable curvature. 

We can obtain as well the coefficient of the logarithmic term of $W$.
Upon integrating $U'(r)$ twice over $r$, we conclude that
\be
{W(r)\over A} = {a_1\over 8\pi}\,\ln r + C_1\,r + C_2.
\ee
Hence we can try to connect with the critical value stated in
(\ref{critW}) as follows.  Let us return to the heat-kernel
representation and perform the divergent integral for $W(r)$,
Eq.~(\ref{Whk}), with a lower cutoff:
\be
{W(r)\over A} = -{a_1\over 8\pi}\int\limits_\e^\infty {dt\over t}\,
e^{-m^2\,t} = -{a_1\over 8\pi}\,\Gamma(0,m^2\,\e) =
-{a_1\over 8\pi}\,[-\ln (m^2\,\e) - \gamma + {{O}(m^2\,\e)}],
\ee
where $\Gamma(0,x)$ is the incomplete gamma function. This agrees with
the previous result. We may interpret the cutoff in
$t$ as an UV cutoff, $\e \sim \Lambda^{-2}$. Reinstating the radius of
curvature we can then split $m^2\,\e$ so that
$$
\ln (m^2\,\e) = \ln r - \ln ({\Lambda^{2}\, R^2}).
$$
The second term agrees with the critical value, Eq.~(\ref{critW}),
but the first term diverges as $r\rightarrow 0$. In fact, this limit
is not meaningful since the logarithmic term arises in the asymptotic
expansion for large $r$. Neither is it meaningful to utilize the
asymptotic expansion of $K(t)$, Eq.~(\ref{aK}), to obtain the small-$r$
behaviour: for $r=0$ the integrals for thermodynamic quantities are
determined by the entire range of $t$ and not just by its asymptotic
behaviour, whether it be for small or large $t$. Nevertheless, 
Eq.~(\ref{aK}) can still be utilized to determine the form of the
divergence of $W(0)$ at $t=0$. Indeed,
\be
W[0,\Lambda] = -{A\over2}\int\limits_{\Lambda^{-2}}^\infty {dt\over t}\,K(t)
\ee
implies that
\be
\Lambda{\partial W[0,\Lambda]\over\partial \Lambda} = 
-A\,K(\Lambda^{-2}) = -{A\over 4\pi} 
\left[{\Lambda^{2}}+ a_1 + O(\Lambda^{-2})\right].
\ee
In the infinite-cutoff limit, $\Lambda\,R \gg 1$, we can discard the
negative powers of $\Lambda$ and this equation is none but the conformal
anomaly equation for a rigid scale transformation (recall that $a_1$
is just the Riemann curvature divided by 3). Integration over
$\Lambda$ yields the desired term, 
$$-{A\,a_1\over 4\pi}\,\ln ({\Lambda\,R})= -{\chi\over 6}\,\ln
({\Lambda\,R}),$$
on a compact surface.

Notice however that $W(0)$ can also be calculated with zeta-function
regularization and that it yields a finite result \cite{D'H-Ph}.  This
may seem puzzling, for we have then lost track of the conformal
anomaly.  In any event, $W(0)$ is a pure number and its value is to
some extent irrelevant. Nevertheless, in the massive case, zeta-function
regularization also yields a finite $W(r)$, as we shall remark in
the next section.  But then the conformal anomaly can be extracted
from the asymptotics of $W(r)$ for large $r$. Moreover, with any type
of regularization, the asymptotics of $S(r)$ for large $r$ provides
the conformal anomaly. This conclusion is not particularly interesting
for the massive bosonic field theory, the only one considered here,
but it may be very interesting for interacting theories.

We have concluded that the coefficient of the logarithmic term in the
entropy must be the opposite of the conformal anomaly, that is,
$\chi/12$.  While we had obtained $-1/3$ in Eq.~(\ref{Ssa}) now we get
$\chi/12 = 1/6$. This discrepancy stems from having suppressed the
zero mode in the calculation of $U(r)$ for the sphere. This
contributes $1/2$ in any compact surface and $1/2 - 1/3 = 1/6$. We may
make a little disgression here and recall that the calculation of the
small-$t$ behaviour of tensor Laplacians on compact Riemann surfaces
provides a proof of the Riemann-Roch theorem \cite{Alva,AG-Nel}. In
the case of the scalar Laplacian, 
$\Delta = {\bar{\partial}}^{\dag}\,{\bar\partial}$, 
this theorem states that
\be
{\cal I}({\bar\partial}) = \dim \Ker {\bar{\partial}} - \dim \Ker
{\bar{\partial}}^{\dag} = 1-g,
\ee
where the adjoint $\bar{\partial}^{\dag} = \nabla_z^1$ is the
covariant derivative on forms. Then $\Ker \nabla_z^1$ is the space of
Abelian differentials, with dimension $g$ and $\Ker \bar{\partial}$ is
the space of holomorphic functions, that is, constants, with dimension
1 (the zero mode). Their difference is $1-g = \chi/2$. According to
our discussion on the presence of logarithmic terms in $r$, we can
interpret the zero mode as the logarithmic term for $r\rightarrow 0$
whereas the logarithmic terms found in the asymptotics $r\rightarrow
\infty$ are related to the existence of non-trivial boundary
conditions and hence with the existence of Abelian differentials. They
substract from 1 the right number $g$ such that the difference is
proportional to the Euler-Poincar\'e number. Indeed, we observe that, e.g.,
for the torus the zero-mode term $\ln\left[1- e^{-L\,m}\right] \approx
\ln (m\,L)$ as $m\rightarrow 0$, while it vanishes exponentially
in the opposite limit, $m\rightarrow \infty$. This is in accord with
the torus being flat, so that $\chi = 0$.

\section{Thermodynamic quantities for compact Riemann surfaces of higher genus}

A compact Riemann surface of $g > 1$ can be characterized by its
fundamental group.  When this surface is represented in its covering
space, the hyperbolic plane, it gives rise to a tesselation of it,
whose tiles are identified by a discrete group of motions isomorphic
to the fundamental group. Since the total group of motions of the
hyperbolic plane is $SL(2,\IR)$, that group is one of its discrete
non-commutative subgroups, which are called Fuchsian groups. This
construction is analogous to the construction of the torus by
factoring the plane by a lattice $\Omega$, where the fundamental group
is $\IZ \times \IZ$. We have seen that there is an expression for the
heat kernel of the torus as the kernel in the plane plus a series of
powers of $\exp (-1/t)$ (\ref{Jac}). There is a non-commutative
analogue for $g > 1$, namely, the Selberg trace formula \cite{McKean},
\be
K(t) =  \frac{e^{t/4}}{2\pi}
\int\limits_0^\infty d\nu\, \nu\,{\tanh (\pi\nu)}\,{e^{-\nu^2\,t}}
+ {1\over2\,A}\sum_{n=1}^\infty \sum_{\{\gamma\}} 
\frac{l_\gamma}{\sinh (n\,l_\gamma/2)}\,\frac{e^{-t/4}}{(4\pi t)^{1/2}}\,
\exp [-{(n\,l_\gamma)^2\over 4\,t}].
\label{Selb}
\ee 
In this formula $\{\gamma\}$ are the primitive conjugacy classes of
the Fuchsian group and $l_\gamma$ is the length of the shortest
geodesic along the corresponding non-contractible loop. The first term
is just the integral representation of the heat kernel in the
hyperbolic plane $H$ (\ref{iK}). It admits a different representation,
more useful for computations, derived from the form of the
corresponding Green function \cite{McKean,Alva,D'H-Ph},
$$
K_H(t) = \int\limits_0^\infty db\, {b\,e^{-b^2/(4t)} \over \sinh{(b/2)}}.
$$

Although the Selberg trace formula is a much more complicated formula
than the Abelian one (\ref{Jac}), it has the same structure; namely,
it is a sum of the part corresponding to the infinite surface, now the
hyperbolic plane, plus $O[\exp (-1/t)]$ corrections due to the
boundary conditions.  As well as for the torus, these corrections give
rise to exponentially vanishing terms which do not appear in the
asymptotic expansion of the entropy around $r=\infty$. Therefore, 
all the higher genus Riemann surfaces share the same asymptotic 
expansion in $r$. 

Formula (\ref{Selb}) has been widely used to establish the modular
dependence of partition functions in string theory
\cite{D'H-Ph,AG-Nel}.  The partition functions in string theory are
those of critical theories and Selberg's formula results in
generalizations of the Dedekind function of the torus to $g > 1$; they
are called Selberg's zeta functions \cite{McKean}.  Although we are
interested here in the massive case, the procedure to calculate the
critical $W$ \cite{McKean,D'H-Ph} actually applies to the non-critical
one as well. It consists of a part corresponding to the hyperbolic
plane, hence common for any homogeneous surface with $g > 1$, and a
part in terms of Selberg's zeta function, which accounts for the
topology of the surface:
\be
W(r) =-{1\over 2} \,\z_r'(0) - 
({4\,r + 1})^{-1/4}\,\ln Z_S\!\left(\sqrt{r + {1\over4}} + {1\over2}\right).
\ee
The zeta-function of the hyperbolic plane is 
\bea
\z_r(x) = {A \over \Gamma(x)}
\int\limits_0^\infty {dt\over t}\,
t^x\,\frac{e^{-(r+1/4)\,t}}{(4\pi\,t)^{3/2}}
\int\limits_0^\infty db\, {b\,e^{-b^2/(4t)} \over \sinh{(b/2)}} = \\
{A\over 4\,{\pi}^{3/2}\,\Gamma(x)}\,(1+ 4\,r)^{3/4 - x/2}
\int\limits_0^\infty db\, {b^{x- 1/2} \over \sinh {(b/2)}}\,
K_{3/2 - x}\!\left(b\,\sqrt{r + {1\over4}}\right).
\eea
Notice that it is a meromorphic function of $x$ with one single pole
at $x=1$, as long as $|r| < 1/4$. In the critical case the common part
can be calculated exactly to yield
\be
\z_0'(0) = 2\,(g-1)\left[-\ln(2\pi) + {1\over 2} - 4\,\z'(-1)\right].
\ee
Selberg's zeta functions is defined as 
\be
Z_S(x) = \prod_\gamma \prod_{p=0}^\infty 
\left(1 - e^{-(x + p)\,l_\gamma}\right).
\ee
The function $\z_r(x)$, as a regularization of $W$ on the hyperbolic
plane, must lead to the asymptotic expansion of the relative entropy
provided by Eqs.~(\ref{Ksa}) and (\ref{entK}). The Selberg zeta function 
leads to the asymptotically vanishing corrections.

\section{Conclusions}

We have seen that the general structure of the entropy of the free
massive bosonic field theory on compact homogeneous Riemann surfaces
consists of a part corresponding to the maximally-symmetric surface,
namely, to the sphere, the plane or the hyperbolic plane, and a part
due to the boundary conditions. The first part can be expressed as a
complicated function, analytic in $r \in [0,\infty)$.  Furthermore, it
has an asymptotic expansion around $r=\infty$, which is fully
computable. The second part embodies the topology and is more delicate
to treat, but it vanishes exponentially as $r$ grows and therefore
does not appear in the asymptotic expansion. The behaviour of the
entropy for small $r$---the critical limit---is also calculable, in
terms of a convergent series.  It is strongly dependent on the global
parameters defining the topological nature of the surface.  Indeed,
one can observe, for example, that the series for the sphere
(\ref{SsT}) has nothing to do with the one for the cylinder, obtained
in \cite{I}. The $k$th term of the small-$r$ expansion is easily seen
to be proportional to $\sum_n \left(-\gamma_n\right)^{-k}$, beginning
with $k=2$.  But for $g > 1$ these sums cannot be computed
analytically, because the Laplacian eigenvalues are not available.

The monotonic character of the entropy can be checked in our
calculations. For large $m$ the entropy tends to $m^2/(8\pi)$,
independently of the topology of the surface, which is the value for
the plane. However, the topology lets itself be felt in the subleading
term, proportional to $\ln m$, which is actually related to the
conformal anomaly. These two terms are the only ones divergent as
$m\rightarrow\infty$. There is also a constant term in the asymptotic
expansion, which cannot be determined exactly, however.  The three
terms together already provide a good approximation down to $m\,L \sim
1$.  As we decrease $m$ we approach criticality and the entropy
becomes very sensitive to large scale peculiarities of the surface, as
already remarked, but one can check that it always remains monotonic.

We may wonder how much of the above can be generalized to higher
dimensions or to interacting field theories.  The spectrum of the
Laplacian and the heat kernel for homogeneous spaces are well known
\cite{Campo}. In fact, the heat kernel is simpler in odd-dimensional
homogeneous spaces than in even-dimensional ones, so the formulas for
the three-dimensional sphere or hyperbolic space turn out to be
simpler as well. As regards massive interacting theories, an
asymptotic expansion for large $m$ must exist in general and,
moreover, the leading and subleading terms can also be studied.  The
leading term is always proportional to $m^2$ for dimensional reasons,
and its coefficient positive. To determine this coefficient one can
use the thermodynamic-Bethe-ansatz computation of the {\em universal
  bulk term} of the free energy, as discussed before \cite{I}. The
subleading term is likely to be related to the conformal anomaly,
$c\,(\chi/6)\,\ln m$, where $c$ is the central charge of the conformal
field theory for $m=0$.  This may provide a new way to find the
central charge of a model.  The small-$m$ behaviour can be studied
with conformal perturbation theory, see Ref.~\cite{I}.  However, it
shall crucially depend on the nature of the surface, as well as on the
particular field theory. This perturbation theory is presumably
convergent for {\em strongly relevant} perturbations, like that of the
free bosonic field theory.

\section*{Acknowledgments}

I acknowledge partial support under Grant PB96-0887.  I thank Alvaro
Dom\'{\i}nguez for conversations and Matt Visser for a careful reading
of a preliminary version of this manuscript.

\appendix

\section*{Appendix}

\subsection*{Substracted energy for the sphere}

We here perform the computation of the substracted energy for the sphere. 
\be
U(r) = -{r\over2} \left[\sum_{l=1}^{\infty}\frac{1}{[l\,(l+1)+r]\,(l+1)} + 
\sum_{l=1}^{\infty}\frac{1}{[l\,(l+1)+r]\,l} \right];
\ee
taking into account that
\bea
\sum_{l=1}^{\infty}\frac{1}{[l\,(l+1)+r]\,l} = \nonumber\\
{1 \over 2\,r}\left[2\,\gamma + 
  \left( 1 + {\frac{1}{{\sqrt{1 - 4\,r}}}} \right) \,
   \psi(
    {\frac{3 - {\sqrt{1 - 4\,r}}}{2}}) + 
  \left( 1 - {\frac{1}{{\sqrt{1 - 4\,r}}}} \right) \,
   \psi({\frac{3 + {\sqrt{1 - 4\,r}}}{2}})\right],
\eea
\bea
\sum_{l=1}^{\infty}\frac{1}{[l\,(l+1)+r]\,(l+1)} = \nonumber\\
{\frac{-\left( \left( 1 - \gamma \right) 
        \left( 1 - 4\,r - {\sqrt{1 - 4\,r}} \right)  \right)  - 
     \left({\sqrt{1 - 4\,r}} + 2\,r  -1 \right)
      \psi(
        {\frac{1-{\sqrt{1 - 4\,r}}}{2}}) - 
     2\,r\,\psi(
       {\frac{1 + {\sqrt{1 - 4\,r}}}{2}})}{\left( -1 + 
       {\sqrt{1 - 4\,r}} \right) \,{\sqrt{1 - 4\,r}}\,r}},
\eea
we have that 
\bea
U(r) = {1 \over 4}
\left\{-2\,\gamma - 
  \left( 1 + {\frac{1}{{\sqrt{1 - 4\,r}}}} \right) 
   \psi({\frac{3 - {\sqrt{1 - 4\,r}}}{2}}) - 
   \left( 1 - {\frac{1}{{\sqrt{1 - 4\,r}}}} \right) 
   \psi({\frac{3 + {\sqrt{1 - 4\,r}}}{2}}) \right. \nonumber\\ \left. 
  +{\frac{2\left( \left( 1 - \gamma \right)
         \left( 1 - 4\,r - {\sqrt{1 - 4\,r}} \right)  + 
        \left({\sqrt{1 - 4\,r}} + 2\,r -1 \right) 
         \psi(
          {\frac{1 - {\sqrt{1 - 4\,r}}}{2}}) + 
        2\,r\,\psi(
          {\frac{1 + {\sqrt{1 - 4\,r}}}{2}}) \right) }{
      \left( -1 + {\sqrt{1 - 4\,r}} \right) {\sqrt{1 - 4\,r}}}
    } \right\}.
\eea
The apparent pole at $r = 0$ must cancel and a careful analysis shows
that it does; the pole at $r = 1/4$ cancels as well.  Notice that when
$r > 1/4$ the argument of the digamma functions becomes complex.
Nevertheless, $U(r)$ remains real and is an analytic function of $r$
at $r = 1/4$.

\subsection*{$W(r)$ for the torus}

Let us $L$ and $M$ denote the periods in the horizontal and vertical
directions, respectively. Then
\begin{equation}
W[m] \equiv -\ln Z[m] =
{1\over2}\sum_{l,n=-\infty}^{\infty}
\ln\left[{\left({2\pi l\over L}\right)^2 + 
\left({2\pi n\over M}\right)^2 + m^2}\right],
\label{Wt}
\end{equation}
and 
\be
U(r) := {dW\over dr} = {1\over2}\sum_{l,n=-\infty}^{\infty} 
\frac{1}{(l/L)^2+(n/M)^2+r},
\ee
where $r := ({m/2\pi})^2$. To work out this sum we can use the known 
expansion of the hyperbolic cotangent in simple fractions \cite{GrRy},
\be
\coth (\pi x) = \frac{x}{\pi}\sum_{n=-\infty}^{\infty}\frac{1}{n^2+x^2}.
\ee
Hence,
\be
U(r) = \frac{1}{2}\sum_{l=-\infty}^{\infty}\frac{\pi\,M}{\sqrt{(l/L)^2+r}}\,
\coth ({\pi}\,M\,{\sqrt{(l/L)^2+r}}).    \label{Ut}
\ee

Now we can obtain $W(r)$ by integration. Notice, however, that the
series (\ref{Ut}) is divergent, so term by term integration is not
warranted. However, the series for $U(r) - U(0)$ is convergent and one
can apply it to it. The ensuing series represents $W(r) - U(0)\,r$
minus a UV quadratically divergent constant. Of course, $U(0)\,r$ is
the UV logarithmic divergence of $W(r)$. Since the UV divergences of
$W(r)$ can be easily segregated, we can proceed with the integration
term by term without further concern:
$$
\frac{M}{2}\int \frac{\pi\,dr}{\sqrt{(l/L)^2+r}}\,
\coth ({\pi}\,M\,{\sqrt{(l/L)^2+r}})  = 
\frac{M}{2}\int d\e\,\coth (\frac{M}{2}\,\e) = \ln\sinh(\frac{M}{2}\,\e),
$$
where we have introduced the one-boson energies 
$\e(l) = \sqrt{(2\pi\,l/L)^2+m^2}$. Finally,
\be
W(r) = \sum_{l=-\infty}^{\infty}\ln\sinh(\frac{M}{2}\,\e(l))= 
\frac{M}{2}\sum_{l=-\infty}^{\infty}\e(l)+
\sum_{l=-\infty}^{\infty}\ln\left[1- e^{-M\,\epsilon(l)}\right] + C,
\ee
where $C=-\ln 2\sum_{l=-\infty}^{\infty}1$ is an irrelevant divergent
constant. The other divergences appear in the first term and are of
the form $C_1 + C_2\,r$, as already remarked.

\end{document}